\documentstyle[psfig,epsf,conf-X,10pt]{article}
\begin{document} 
\small
\heading{%
%
Closing the gap: an X-ray selected sample of clusters of galaxies
behind the Galactic plane
}
\par\medskip\noindent
\author{%
Harald Ebeling, Christopher Mullis, Brent Tully
}
\address{%
Institute for Astronomy, 2680 Woodlawn Drive, 96822 Honolulu, HI, USA
}

\begin{abstract}
We discuss the design and current status of the CIZA survey, the first
systematic X-ray search for clusters of galaxies in the Galactic plane
region. So far, we have compiled a sample of more than 70 X-ray
selected clusters at $|b|<20^{\circ}$, 80\% of which were previously
unknown. Upon its completion the CIZA cluster catalogue will
complement the existing galaxy surveys in the Galactic plane region
and allow a fresh look at large-scale structure and local streaming
motions.
\end{abstract}
\section{Introduction}
Clusters of galaxies are X-ray bright to the extent that the ROSAT
All-Sky Survey (RASS) allows sizeable, statistically complete cluster
samples to be compiled \cite{ebe1,ebe2,deg}. Large-scale structure
(LSS) studies using clusters as tracers do not duplicate, but rather
complement those using galaxies because clusters mark the locations of
the deepest potential wells whereas galaxies probe primarily the
low-density field. To date, most dynamical analyses of large-scale
flows have been compared with the IRAS density fields
\cite{dekel,davis}. However, the distribution of rich clusters is
significantly different from that of IRAS-selected galaxies
\cite{kaiser,fisher} as the latter are mostly spirals. There is
evidence from dynamical modeling that mass congregates in
clusters with much higher $M/L$ values than associated with field
galaxies \cite{brent1}.

Historically, optical searches for clusters of galaxies, and thus also
the resulting LSS studies, were forced to avoid a wide band of the sky
centered on the Galactic plane because of severe extinction and
stellar obscuration at $|b|<20^{\circ}$. With the advent of X-ray
astronomy, this restriction is greatly relaxed. Rather than dust
extinction and stellar confusion, it is now the X-ray absorbing
equivalent Hydrogen column density, $n_{\rm H}$, that is the limiting
factor. As shown in Figure 1, $n_{\rm H}$ rises only slowly toward the
Galactic plane allowing an X-ray cluster survey to penetrate the
plane to much lower latitude.

\section{CIZA: closing the gap}

In an attempt to overcome the limitations of existing cluster samples,
we have initiated a program aimed at the construction of a statistical
sample of galaxy clusters in what used to be the zone of
avoidance. Our survey is based on X-ray sources detected in the RASS
as listed in the ROSAT Bright Source Catalog \cite{voges}. In a first
phase of our project which focuses on the X-ray brightest clusters we
apply three selection criteria: $|b|<20^{\circ}$, $f_{\rm X}\ge
3\times10^{-12}$ erg cm$^{-2}$ s$^{-1}$, and a spectral hardness ratio
cut to discriminate against softer, non-cluster X-ray sources.

Cross-correlations with databases of Galactic and extragalactic
objects as well as follow-up observations with the University of
Hawai`i's 2.2m telescope have, so far, resulted in 73
spectroscopically confirmed galaxy clusters at $0.02\le z\le 0.34$;
only 15 of which were previously known (see Figure 1).  Highlights of
the survey so far include the discovery of a distant, extremely X-ray
luminous cluster which acts as a gravitational lens, and, at the
opposite end of the redshift scale, the discovery of a cluster at $z=
0.022$ at $|b|=0.3^{\circ}$ (i.e., in the mid-plane of the Milky Way)
which is very likely part of the Perseus-Pegasus complex.

\begin{figure}
\mbox{\epsfxsize=12cm \hspace*{-3.5mm} \epsffile{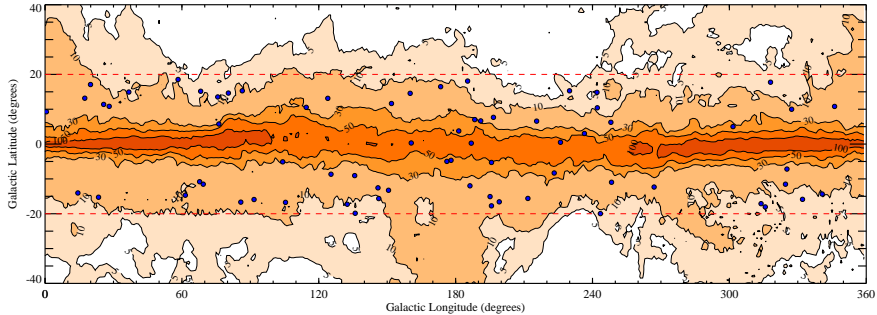}}
\caption[]{Contours of the Galactic $n_{\rm H}$ column density at
$|b|<40^{\circ}$ in units of $10^{20}$ cm$^{-2}$. The solid circles
show the locations of the clusters discovered and spectroscopically
confirmed by our survey prior to April 1999.
}
\end{figure}
\vfill
\end{document}